\def\half{\hbox{$1\over2$}}

\def\sib{\bar\sigma}

\def\kab{\bar\kappa}

\def\rhb{\bar\rho}
\def\pib{\bar\pi}

\def\tab{\bar\tau}

\def\kap{\kappa'}
\def\sip{\sigma'}
\def\rhp{\rho'}
\def\tap{\tau'}

\font\ec=ecrm0800 at 12pt
\def\th{\hbox{\ec\char'336}}
\def\edth{\hbox{\ec\char'360}}
\def\thp{\hbox{\ec\char'336}'}
\def\edthp{\hbox{\ec\char'360}'}

\def\thpt{\tilde{\th}'}
\def\edtht{\tilde{\edth}}
\def\edthpt{\tilde{\edth}'}

\def\pheq{\phantom{=}}

\def\deg#1{{#1}^\circ}

\def\mb{\bar m}

\documentclass[12pt,epsf]{iopart}
\usepackage{iopams}

\begin{document}
\title[Radiation Gauges in type II spacetimes]{On the Existence of
Radiation Gauges in Petrov type II spacetimes}
\author{Larry R. Price\dag, Karthik Shankar\ddag, Bernard
F. Whiting\P}
\address{Department of Physics, University of Florida, PO Box 118440,
Gainesville, FL 32611, USA}
\ead{\dag price@phys.ufl.edu,\ddag karthik@phys.ufl.edu,\P
bernard@phys.ufl.edu}
\begin{abstract}
The radiation gauges used by Chrzanowski (his IRG/ORG) for metric
reconstruction in the Kerr spacetime seem to be over-specified.  Their
specification consists of five conditions: four, which we treat here
as valid gauge conditions, plus an additional condition on the trace of the
metric perturbation.  In this work, we utilize a newly developed form of the
perturbed Einstein equations to establish a condition --- on a particular
tetrad component of the stress-energy tensor --- under which the full IRG/ORG
can be  imposed. Using gauge freedom, we are able to impose the full
IRG for Petrov type II and type D backgrounds, using a different
tetrad for each case.  As a specific example, we work through the process of
imposing the IRG in a Schwarzschild background, using a more traditional
approach.  Implications for metric reconstruction using the Teukolsky curvature
perturbations in type D spacetimes are briefly discussed.
\end{abstract}
\maketitle

\section{Introduction}
The Regge-Wheeler\cite{Regge:1957td} (RW) approach to perturbations of the
Schwarzschild spacetime is usually understood to lead, by direct integration,
to  perturbations for all parts of the metric in terms of gauge invariant
quantities\cite{Moncrief:1974}.  In fact, it has been evident for a long
time\cite{PhysRevD.5.2439} that the RW variable actually represents part of the
perturbation of the Weyl curvature (namely, ${\rm
Im}{(\Psi_2)}$\cite{Nolan:2004hc}; see \cite{Jezierski:1998ui} for a clear
demonstration of this and \cite{Whiting:2005hr} for further discussion).  These
contrasting perspectives are reconciled by the exceptional fact that the
geometrical symmetries of the Schwarzschild spacetime permit an analysis
virtually transparent to both angular and time derivatives.

For the perturbations of the Kerr spacetime, the situation is completely
different.  Instead of the RW equation, we have the Teukolsky equation for the
gauge (and tetrad) invariant parts ($\Psi_0$ and $\Psi_4$) of the perturbed
Weyl curvature.  To date, metric reconstruction\cite{Ori:2002uv, Lousto:2002em}
can then be performed by a (Hertz) potential method championed by
Chrzanowski\cite{Chrz:1975wv}, but only fully in vacuum, and even so, only in a
special class of over-specified gauges referred to as ``radiation gauges''.  We
expect that Hertz potential methods are set to play an increasingly key r\^ole
as we pursue deeper studies of perturbations of Petrov type D spacetimes.
However, neither Chrzanowski's analysis, nor the more general analyses of Cohen
and Kegeles\cite{CKI,Kegeles:1979an} and Stewart\cite{Stewart} for
perturbations of Petrov type II spacetimes, spells out the precise
circumstances in which radiation gauges are able to exist.\footnote[1]{However,
the constructive procedure of Stewart does go a long way in this direction.}
The purpose of this paper is to address and dispel this concern, by specifying
exactly when a radiation gauge may be imposed.

The Petrov classification refers to the properties of the (always null)
eigenvectors of the Weyl tensor, referred to as principle null directions
(PNDs).  In a type II spacetime, one of these PNDs is repeated.  It may be
either the ingoing null vector $l^{a}$ or the outgoing null vector $n^{a}$.  In
a type D spacetime, two of the PNDs are repeated, namely, both $l^{a}$ and
$n^{a}$.  Radiation gauges have been defined\cite{Chrz:1975wv} with respect to
either one of these PNDs. In a suitable spacetime, with metric $g_{ab}$ and for
metric
perturbation $h_{ab}$,\footnote{Throughout, we have consistently used the
perturbed metric to be $g_{ab}-h_{ab}$.} the gauge conditions specified for a
radiation gauge are either:
\begin{description}
\item[ i)] $l^{a}h_{ab}=0$ and $g^{ab}h_{ab}=0$, referred to as
``Ingoing'' (IRG), or \item[ ii)] $n^{a}h_{ab}=0$ and
$g^{ab}h_{ab}=0$, referred to as ``Outgoing'' (ORG).
\end{description}
In each case, since these represent five distinct conditions, it is clear that
radiation gauges cannot be defined in general, but it turns out they can
be prescribed in special circumstances, which we have investigated here.

In general terms, in type D background spacetimes, two radiation gauges are
indeed possible.  In type II background spacetimes, depending on which
principle null direction is repeated, only one or the other of these gauges
would be possible.  In all cases, we find that radiation gauges can normally
actually exist only for a perturbed stress tensor, ${\cal T}_{ab}$, satisfying
${\cal T}_{ab}l^{a}l^b\equiv{\cal T}_{ll}=0$ in the IRG case or ${\cal
T}_{ab}n^{a}n^b\equiv{\cal T}_{nn}=0$ in the ORG case.  This is because, in the
appropriate circumstances, the $ll$- (or $nn$-) component of the perturbed
Einstein equation depends only on the trace of the metric perturbation, which
cannot be zero if the source for that equation is non-zero.  Remarkably, in all
type II background spacetimes (which include type D as a special case), a
non-zero solution to the trace equation without  sources can be fully gauged
away by the use of residual gauge freedom.

The layout of the paper is as follows.  We first introduce a new form of the
perturbed Einstein equations in the Newman-Penrose formalism.  Then, we
describe the radiation gauges in more detail, followed by explanations of how
they are set up in type II and type D background spacetimes, respectively.
This requires us to examine the perturbed Einstein tensor to understand fully
the
implications of attempting to impose a radiation gauge.  We illustrate the
residual gauge freedom and the condition for it to remove the trace of the
perturbed metric. Next, we demonstrate the implication of our analysis in the
Schwarzschild spacetime.  Finally we contrast the construction of Hertz
potentials and the construction of radiation gauges.

\section{A new form of the perturbed Einstein equations}
We choose to work with a formulation of the perturbed Einstein equations that
makes explicit use of the modified Newman-Penrose\cite{NP} (NP) formalism of
Geroch, Held and Penrose\cite{GHP} (GHP).  For a detailed explanation of the
GHP formalism see also \cite{PENRIND}.  The starting point is to take a
complex null tetrad $\{l^a,n^a,m^a,\mb^a\}$, where an overbar denotes the
complex conjugate, normalized so that\footnote{The conventions displayed in
\eref{tetrad:def} and \eref{metric:def}  with signature [+,-,-,-] are
characteristic of the NP formalism.}
\begin{equation}
l^an_a = -m^a\mb_a = 1.\label{tetrad:def}
\end{equation}
Then, the spacetime metric has the following expression:
\begin{equation}
g_{ab}=2l_{(a}n_{b)} - 2 m_{(a}\mb_{b)},\label{metric:def}
\end{equation}
in which round brackets, $()$, around indices denote symmetrization. Note that
the metric is invariant under the transformation
\begin{equation}
\eqalign{l^a   &\to \zeta{\overline\zeta}l^a, \\
         m^a   &\to \zeta{\overline\zeta}^{-1}m^a,}
\qquad
\eqalign{n^a   &\to \zeta^{-1}{\overline\zeta}^{-1}n^a,\\
         \mb^a &\to \zeta^{-1}{\overline\zeta}\mb^a,}
         \label{eqn:typetrans}
\end{equation}
for some complex number $\zeta$.  A key feature of the GHP formalism is that
all objects of interest transform homogeneously under \eref{eqn:typetrans}.  A
quantity, $\chi$, is said to be of type $\{p,q\}$ if under \eref{eqn:typetrans}
it transforms as $\chi\to\zeta^p{\bar\zeta}^q\chi$.  Alternatively, it useful
to speak of the spin-weight $s=(p-q)/2$ and boost-weight $b=(p+q)/2$ of $\chi$.
 A table
listing the types of the fundamental GHP quantities can be found in
\cite{GHP}.

We can express the metric perturbation, $h_{ab}$, in terms of the tetrad
vectors according to
\begin{eqnarray}
\eqalign{h_{ab} &={} h_{nn}l_al_b - 2h_{n\mb}l_{(a}m_{b)} -
2h_{nm}l_{(a}\mb_{b)} + 2h_{ln}l_{(a}n_{b)}\\
&\pheq +h_{ll}n_an_b - 2h_{l\mb}n_{(a}m_{b)} - 2h_{lm}n_{(a}\mb_{b)}\\
&\pheq +h_{mm}\mb_a\mb_b + 2h_{m\mb}m_{(a}\mb_{b)} + h_{\mb\mb}m_am_b,}
\end{eqnarray}
where $h_{ll}=h_{ab}l^al^b$, $h_{lm}=h_{ab}l^am^b$ and so on, are the tetrad
components of the metric perturbation.  Note that the GHP type of each
component of the metric perturbation is inherited from the tetrad vectors, e.g.
$h_{ll}$ has type $\{2,2\}$, $h_{n\mb}$ has type $\{-2,0\}$, etc.  The
perturbed Einstein tensor, ${\cal E}_{ab}$, is then computed via
\begin{equation}
\fl {\cal E}_{ab}=-\half\Theta^c\Theta_c h_{ab} - \half\Theta_a\Theta_b
h^c{}_c + \Theta^c\Theta_{(a} h_{b)c} + \half
g_{ab}(\Theta^c\Theta_c h^d{}_d - \Theta^c\Theta^d
h_{cd})\label{eqn:GFEE},
\end{equation}
where
\begin{equation}
\Theta_a = l_a\thp + n_a\th - m_a\edthp - \mb_a\edth,\label{eqn:GHPD}
\end{equation}
is just the covariant derivative expressed in GHP language.  We can use the
expression in (\ref{eqn:GHPD}) to define the GHP derivatives `thorn'
($\th=l^a\Theta_a$), `edth' ($\edth=m^a\Theta_a$) and their `primes'
($\thp=n^a\Theta_a$ and $\edthp=\mb^a\Theta_a$).  These derivative
operators also inherit their type from the corresponding tetrad vectors:
\begin{equation}
\eqalign{\th   &:\lbrace 1,1\rbrace,\\
         \edth  &:\lbrace 1,-1\rbrace,}
\qquad
\eqalign{\thp  &:\lbrace -1,-1\rbrace,\\
         \edthp &:\lbrace -1,1\rbrace.}
\end{equation}
It is sometimes useful to think of $\th$ ($\thp$) and $\edth$
($\edthp$) as boost and spin raising (lowering) operators, respectively.  With
these conventions, the perturbed Einstein equations are given in general as
\begin{equation}
{\cal E}_{ab} = 8\pi{\cal T}_{ab},\label{eqn:pEEs}
\end{equation}
where ${\cal T}_{ab}$ is the stress-energy tensor source for the perturbation.

The form we use for the perturbed Einstein equations offers several advantages.
 First, it allows us to deal with perturbations of an entire class of
spacetimes at once, say of Petrov type II or type D.  Furthermore, this level
of generality comes at no expense in terms of the complexity of the equations.
See, for example, \ref{App:typeII} which lists the components of the perturbed
Einstein tensor for an
arbitrary type II background.  Additionally, the requirement that every term in
an expression have the same spin- and boost- weight provides both a useful
check on the equations and new insight into the structure of the perturbed
Einstein equations.  All of this comes with the added benefit of having a
simple and straightforward way of dealing with gauge freedom, which has proved
to be crucial for describing metric
perturbations.  We find this form of the
equations provides a powerful new tool for exploring metric perturbations.

\section{The Radiation Gauges}
The ingoing radiation gauge (IRG) is a crucial ingredient for the
reconstruction of metric perturbations of Petrov type D spacetimes from
curvature perturbations. They first appear, unexplained, in the work of Cohen
and Kegeles
\cite{CKI} (for perturbations of Petrov type II spacetimes) and Chrzanowski
\cite{Chrz:1975wv} (who considered perturbations of Petrov type D spacetimes),
but the work that comes closest to our contribution in describing their origin
is that of Stewart \cite{Stewart}, again for the more general case of type II
spacetimes.

In type II background spacetimes, the IRG is defined by the conditions
\numparts
\begin{eqnarray}
l^ah_{ab} &=& 0,\label{eqn:IRG}\\
g^{ab}h_{ab} &=& 0,\label{eqn:IRGtr}
\end{eqnarray}
\endnumparts
where $l^a$ is aligned with the repeated PND of the background Weyl tensor.  If
$n^{a}$ rather than $l^{a}$ is a repeated PND, we instead define the outgoing
radiation gauge (ORG) by
\numparts
\begin{eqnarray}
n^ah_{ab} &=& 0,\\
g^{ab}h_{ab} &=& 0.\label{eqn:ORG}
\end{eqnarray}
\endnumparts
In type II background spacetimes, only one or the other of these options exists
(IRG or ORG), whereas in  Petrov type D background spacetimes, there is the
possibility of defining both gauges.  For the remainder of this work we focus
on the IRG.  Results for the ORG can be obtained by making the replacement $l^a
\leftrightarrow n^a$.

Equations (\ref{eqn:IRG}-$b$) translate into algebraic conditions on the
components of the metric perturbation.  We will refer to the four conditions in
(\ref{eqn:IRG}) as the $l\!\cdot\!h$ gauge conditions.\footnote{Recently, when
applied specifically to the Schwarzschild spacetime, these conditions were
given a geometrical interpretation, and referred to as {\it light-cone gauge
conditions}\cite{Preston:2006ze}, though they are not the conditions originally
introduced for gravitation with that name\cite{Scherk:1974zm}.  It may well be
that this description is suitable more generally, although presumably without
the specific geometrical interpretation of \cite{Preston:2006ze}.} In terms of
the tetrad components of the metric perturbation, these gauge conditions read:
\begin{equation}
{h_{ll}= 0,\qquad
h_{ln}= 0,\qquad
h_{lm}= 0,\qquad
h_{l\mb}= 0.}\label{eqn:gcomps}
\end{equation}
The condition in \eref{eqn:IRGtr} will be referred to as the trace condition
and can be expressed in terms of the components of the metric perturbation as
$h_{ln}-h_{m\mb} =0,$ which, when (\ref{eqn:gcomps}) is imposed, simply reads
\begin{equation}
h_{m\mb} = 0.\label{eqn:tcomps}
\end{equation}
Because the IRG constitutes a total of five conditions on the metric
perturbation, instead of the four one might expect for a gauge condition, it is
necessary to ensure that the extra condition does not interfere with any
physical degree of freedom in the problem, such as one coming from a source.
The importance of this consideration can be seen immediately from
\eref{eqn:Ell_app} of \ref{App:typeII}, in which every term would be removed by
\eref{eqn:gcomps} and \eref{eqn:tcomps}, rendering \eref{eqn:Ell_app}
inoperable
whenever it has a non-zero source.  In the next section we will look to the
perturbed Einstein equations to determine the circumstances under which we can
safely impose all five of the conditions that constitute the IRG.

It is useful to note the similarity between the full IRG, (\ref{eqn:IRG}-$b$),
and the more commonly known transverse traceless (TT) gauge defined by
\begin{equation}
{\nabla^a h_{ab}= 0,\qquad
g^{ab}h_{ab}= 0,}\label{TTgauge}
\end{equation}
which, at a first glance, also appears to be over-specified.  In fact, the TT
gauge exists for any vacuum perturbation of an arbitrary, globally hyperbolic,
vacuum solution\cite{Wald:book}, because imposing the differential part of the
gauge does not exhaust all of the available gauge freedom.  Interestingly
enough, Stewart's analysis in terms of Hertz potentials\cite{Stewart} begins by
considering a metric perturbation in the TT gauge.  However, in order to
construct the curved space analogue of a Hertz potential,  he is compelled to
perform a transformation that destroys \eref{TTgauge} and instead yields a
metric perturbation in the IRG.\footnote{In flat space, owing to the fact that
partial derivatives commute, this transformation would actually leave one in
the TT gauge.  See \cite{Stewart} or Appendix C of \cite{Whiting:2005hr} for a
more detailed explanation.}  Furthermore it appears that the restriction to
type II spacetimes is essential for Stewart's analysis. {}From these
observations, we expect radiation gauges to exist under conditions less general
than those required for the existence of the TT gauge.  At the same time, we
should not be surprised that the IRG inherits the feature of {\em residual
gauge freedom}.

Consider a gauge transformation on the metric perturbation generated by a gauge
vector, $\xi_a$.  To create a transformed metric in the $l\!\cdot\!h$ gauge,
the gauge conditions (\ref{eqn:gcomps}) require
\begin{equation}
l^a(h_{ab}-\xi_{(a;b)}) = 0,\label{eqn:MPGT}
\end{equation}
where the semicolon denotes the covariant derivative. In terms of components
this reads
\begin{equation}
\eqalign{2\th\xi_l &= h_{ll},\\
\thp\xi_l + \th\xi_n + (\tau+\tab')\xi_{\mb} + (\tab+\tap)\xi_m &=
h_{ln},\\
(\th+\rhb)\xi_m + (\edth+\tab')\xi_l &= h_{lm},\\
(\th+\rho)\xi_{\mb} + (\edthp+\tap)\xi_l &=
h_{l\mb}.}\label{eqn:GTCOMPS}
\end{equation}
Similarly, for the trace condition (\ref{eqn:tcomps}) to be satisfied by the
gauge
transformed metric, we require
\begin{equation}
\edthp\xi_m + \edth\xi_{\mb} +
(\rhp+\rhb')\xi_l + (\rho+\rhb)\xi_n = h_{m\mb}.\label{eqn:TCOMP}
\end{equation}
Any extra gauge transformation that satisfies $l^a\xi_{(a;b)} = 0$, that is,
solves the homogeneous form of (\ref{eqn:GTCOMPS}), preserves the four
$l\!\cdot\!h$ gauge conditions (\ref{eqn:gcomps}). This is what is meant by
residual gauge freedom.  We will explicitly use this residual gauge freedom to
impose the $\,l\!\cdot\!h\,$ and trace conditions simultaneously, thus
establishing the IRG.  We will find that some gauge freedom still remains, as
explained in section \ref{typeII:RGF}.

Now, we turn our attention to the general case of type II background
spacetimes.

\section{Imposing the IRG in type II}\label{typeII}
In order to show that residual gauge freedom can be used to impose the IRG, we
need to solve for the residual gauge freedom as well as examine any perturbed
Einstein equation that might impede the imposition of the trace condition of
the IRG.  For this, we turn to a coordinate-free integration method develop by
Held.  Rather than give a detailed explanation, we present the basics and refer
the interested reader to the literature for an in-depth
account\cite{held:1974,Stewart:1974}.

The first step is to introduce new derivative operators $\thpt$, $\edtht$ and
$\edthpt=\bar{\edtht}$ such that they commute with $\th$ when acting on
quantities that $\th$ annihilates,\footnote{Such quantities are denoted with
the degree mark, $\deg{}$, as in $\th \deg x=0$.} that is
\begin{equation}
{[\th,\thpt]\deg x=0,\qquad
         [\th,\edtht\phantom{'}]\deg x=0,\qquad
	 [\th,\edthpt]\deg x=0,}
\end{equation}
where $[a,b]$ denotes the commutator between $a$ and $b$.  The explicit form of
the operators is given in \ref{Held}.  The next step, the heart of Held's
method,  is to exploit the GHP equation $\th\rho = \rho^2$, and its complex
conjugate $\th\rhb=\rhb^2$, to express everything as a polynomial in terms of
$\rho$ and $\rhb$, with coefficients that are annihilated by $\th$.  Held's
method is then brought to completion by choosing four independent quantities to
use as coordinates\cite{held:1975,Edgar:1992}.  In this work, we will not take
this extra step. For type II spacetimes (and the accelerating C-metrics), 
this step has not been carried
out, while for all remaining type D spacetimes, 
it has been carried through to
completion\cite{held:1974,Stewart:1974}.

In a spacetime more general than type II, there is no possibility of having a
repeated PND.  When a repeated PND exists, we can appeal to the Goldberg-Sachs
theorem \cite{gsachs} and set $\kappa=\sigma=\Psi_0=\Psi_1=0$ in (A.1-A.7).
Following Held's partial integration of Petrov type II
backgrounds\cite{held:1975}, we also perform a null rotation (keeping $l^a$
fixed, but changing $n^a$) to set $\tau=0$.  As a consequence, it follows from
the GHP equations that $\tap=0$.  Now we are in a position to address the
question of when the full
IRG can be imposed.  First we apply the $l\!\cdot\!h$ gauge conditions
(\ref{eqn:gcomps}) to (A.1-A.7).  While most of the perturbed Einstein
equations depend on several components of the metric perturbation, after
imposing
\eref{eqn:gcomps}, the expression for ${\cal E}_{ll}$ depends only on
$h_{m\mb}$ and the $ll$-component of \eref{eqn:pEEs} simply becomes
\begin{equation}
\{\th(\th-\rho-\rhb) +
2\rho\rhb\}h_{m\mb}\equiv\{(\th-2\rho)(\th+\rho-\rhb)\}h_{m\mb}= 8\pi{\cal
T}_{ll}, \label{eqn:Ell}
\end{equation}
in which the first form indicates that the equation is real, while the second
form and its complex conjugate (which follow from the fact that
$\th\rho=\rho^2$ and $\th\rhb=\rhb^2$) is the one we will use to integrate the
equation below.  If we had not made use of the Goldberg-Sachs theorem, there
would be terms such as $\sigma\rho h_{\mb\mb}$ appearing in \eref{eqn:Ell} and
our argument would not hold.   We immediately see that ${\cal T}_{ll}=0$ is
necessary to satisfy the trace condition \eref{eqn:tcomps}.  Next we turn our
attention to the question
of whether the condition ${\cal E}_{ll}=0$, is sufficient to impose
\eref{eqn:tcomps} using residual gauge freedom.

In order to address this question we will integrate ${\cal E}_{ll}=0$ and the
residual gauge vector, given by the homogeneous form of \eref{eqn:GTCOMPS}.
Full integration of the homogeneous form of \eref{eqn:GTCOMPS} is carried out
in \ref{Held}, but we will work through the integration of ${\cal E}_{ll}=0$
here to illustrate the method.  We begin by rewriting \eref{eqn:Ell}, with the
help of $\th\rho=\rho^2$ and its complex
conjugate, as:
\begin{equation}
\{(\th-2\rho)(\th+\rho-\rhb)\}h_{m\mb}=
\rho^2\th\Big[\frac{\rhb}
{\rho^{3}}\th\Big(\frac{\rho}{\rhb}h_{m\mb}\Big)\Big] = 0.\label{Ell:trace}
\end{equation}
Integrating once gives
\begin{equation}
\th\Big(\frac{\rho}{\rhb}h_{m\mb}\Big)=\deg{b}\frac{\rho^3}{\rhb},
\end{equation}
and another integration leads to
\begin{equation}
h_{m\mb} = \deg{\bar{a}}\frac{\rhb}{\rho} + \half\deg{b}(\rho+\rhb).
\end{equation}
However, $h_{m\mb}$ is, by definition, a real quantity, so we add the complex
conjugate and use $\deg{b}$ to represent a real quantity in the second term.
The final result is that
\begin{equation}
h_{m\mb} = \deg{a}\frac{\rho}{\rhb} + \deg{{\bar a}}\frac{\rhb}{\rho}
+ \deg{b}(\rho + \rhb)\label{eqn:hmmbsoln}.
\end{equation}
Similarly, integration of \eref{eqn:GTCOMPS}, as carried out in \ref{Held},
leads to the following solution for the components of the residual gauge
vector:
\begin{equation}
\eqalign{ \xi_l &= \deg{\xi_l},\\
          \xi_n &= \deg{\xi_n} +
          \half\Big(\frac{1}{\rho}+\frac{1}{\rhb}\Big)
          \thpt\deg{\xi_l} +
          \half(\deg{\Psi_2}\rho +
          \deg{\bar{\Psi}_2}\rhb)\deg{\xi_l},\\
	  \xi_m &= \frac{1}{\rhb}\deg{\xi_m} -\edtht\deg{\xi_l},\\
	  \xi_{\mb} &= \frac{1}{\rho}\deg{\xi_{\mb}}-
          \edthpt\deg{\xi_l},\\
}\label{eqn:type2compts}
\end{equation}
where $\deg{\Psi_2}$ is related to the background curvature via
$\Psi_2=\deg{\Psi_2}\rho^3$.  In order to use this residual gauge freedom to
impose the full IRG, we return to the gauge transformation for $h_{m\mb}$
\eref{eqn:TCOMP} which becomes, after some manipulation (using
(\ref{eqn:T21}-\ref{eqn:T24}) and \eref{eqn:edthcomm}),
\begin{equation}\fl
h_{m\mb} = \frac{\rho}{\rhb}\Big[\edthpt\deg{\xi_m} +
\thpt\deg{\xi_l}\Big] +
\frac{\rhb}{\rho}\Big[\edtht\deg{\xi_{\mb}}  +
\thpt\deg{\xi_l}\Big] +
(\rho+\rhb) [-\half(\edthpt\edtht + \edtht\edthpt - \deg{\rhp} -
{\rhb}^{\prime\circ} )\deg{\xi_l}
+\deg{\xi_n}]. \label{eqn:GENTYPEIISOL}
\end{equation}
In this form it is clear that we can impose the trace condition
\eref{eqn:tcomps} of the full IRG if we choose our gauge vector so that
\begin{equation}
{ \edthpt\deg{\xi_m}+\thpt\deg{\xi_l}=\deg{a},\qquad
	  -\half(\edthpt\edtht+\edtht\edthpt-\deg{\rhp}-\rhb^{\prime\circ}
	  )\deg{\xi_l} + \deg{\xi_n}=
	  \deg{b}.}\label{gauge:IRG}
\end{equation}
We have now shown by construction that the condition ${\cal T}_{ll}=0$ is both
necessary and sufficient for imposing the full IRG in a type II background.  We
turn next to discussing the complete extent of the residual gauge freedom in
more detail.

\subsection{Remaining gauge freedom}\label{typeII:RGF}

Although equations \eref{gauge:IRG} involve three real degrees of freedom
($\deg{a}$ is complex), it turns out that  only two real degrees of gauge
freedom are required to fully remove any solution of \eref{Ell:trace} for the
trace $h_{m\mb}$.  To see this we introduce the following identity:
\begin{equation}
{\rho\over\rhb}-{\rhb\over\rho}=(\rho+\rhb)\Big({1\over\rhb}
-{1\over\rho}\Big)\equiv(\rho+\rhb)\deg{\Omega},\label{eqn:Omega}
\end{equation}
which also defines $\deg{\Omega}$, a quantity annihilated by $\th$.  Then we
can rewrite \eref{eqn:hmmbsoln} as
\begin{equation}
h_{m\mb} = \half(\deg{a}+\deg{\bar a})
\Big(\frac{\rho}{\rhb}+\frac{\rhb}{\rho}\Big) +
[\half(\deg{a}-\deg{\bar a})\deg{\Omega} + \deg{b}](\rho+\rhb).
\label{eqn:hmmbrw}
\end{equation}
In a similar fashion, we can rewrite \eref{eqn:GENTYPEIISOL} as
\begin{equation}
\eqalign{h_{m\mb} &= \Big[\half(\edthpt\deg{\xi_m}  +
\edtht\deg{\xi_{\mb}}) +\thpt\deg{\xi_l}\Big]
\Big(\frac{\rho}{\rhb}+\frac{\rhb}{\rho}\Big)+\Big[\half(\edthpt\deg{\xi_m}  -
\edtht\deg{\xi_{\mb}})\deg{\Omega}\\
&\pheq-\half(\edthpt\edtht + \edtht\edthpt - \deg{\rhp} -
{\rhb}^{\prime\circ} )\deg{\xi_l}
+\deg{\xi_n}\Big](\rho+\rhb),} \label{eqn:gfrw}
\end{equation}
in which each coefficient in big square brackets is purely real.  Now, suppose
we have a particular solution for ${\cal E}_{ll}=0$, i.e., $\deg a$, $\deg{\bar
a}$ and $\deg b$ are fixed, and our task is to solve for the components of the
gauge vector which removes this solution.  By comparing \eref{eqn:hmmbrw} and
\eref{eqn:gfrw} we see that, for any given $\deg{\xi_m}$ and $\deg{\xi_{\mb}}$,
we can fix $\deg{\xi_l}$ (up to a solution of $\thpt\deg{\xi_l} = 0$) via
\begin{equation}
\thpt\deg{\xi_l} = \half(\deg{a}+\deg{\bar a})-\half(\edthpt\deg{\xi_m}  +
\edtht\deg{\xi_{\mb}}), \label{eqn:lfix}
\end{equation}
and we can fix $\deg{\xi_n}$ by setting
\begin{equation}
\fl\deg{\xi_n} = \half(\deg{a}-\deg{\bar a})\deg{\Omega} + \deg{b} +
\half(\edthpt\edtht + \edtht\edthpt - \deg{\rhp} -
{\rhb}^{\prime\circ} )\deg{\xi_l}-\half(\edthpt\deg{\xi_m}  -
\edtht\deg{\xi_{\mb}})\deg{\Omega}, \label{eqn:nfix}
\end{equation}
to completely eliminate the nonzero $h_{m\mb}$, thus imposing the
full IRG while still leaving two completely unconstrained degrees of gauge
freedom, $\deg{\xi_m}$ and $\deg{\xi_{\mb}}$.  Once in the IRG, then
\eref{eqn:lfix} and \eref{eqn:nfix}, with $\deg{a}$, $\deg{\bar a}$ and
$\deg{b}$ set to zero and $\deg{\xi_m}$ and $\deg{\xi_{\mb}}$ arbitrary, give
the remaining components of a gauge vector preserving the IRG.  It is currently
unclear how to take advantage of this remaining gauge freedom to simplify the
analysis of perturbations in the full IRG.

\section{Imposing the IRG in type D}\label{typeD}
Type D background metrics are of considerable theoretical and observational
interest since they include both the Schwarzschild and Kerr black hole
spacetimes.  Kinnersley first obtained all type D metrics by integrating the
Newman-Penrose equations\cite{kinnersley:1195}.  While the results of the
previous section are general enough to encompass the special case of type D
backgrounds, the tetrad choice we made (with $\tau=0$) is incompatible with the
complete integration of the background field equations which is possible in
type D spacetimes\cite{held:1974}.  The complete integration requires that each
of $l^{a}$ and $n^{a}$ be aligned with one of the two PNDs.  In that case we
can exploit the full power of the Goldberg-Sachs theorem and its corollaries to
set $\kappa=\kap=\sigma=\sip=\Psi_0=\Psi_1=\Psi_3=\Psi_4=0$, while maintaining
$\tau\neq 0$ and $\tap\neq 0$.  In this section we repeat the previous
calculation with this different choice of tetrad.

The result of integrating ${\cal E}_{ll}=0$ is the same as in the case of a
type II background, given in \eref{eqn:hmmbsoln}.  The residual gauge vector,
however, now has the following, more complex, form (details of the integration
are given in \ref{Held}):
\begin{equation}
\eqalign{        \xi_l &= \deg{\xi_l}\!,\\
         \xi_n &= \deg{\xi_n} + \half\deg{\Psi} \deg{\xi_l} \rho+
\half\deg{\bar\Psi} \deg{\xi_l} \rhb +
           \deg\tau \deg\tab \deg{\xi_l} \rho\rhb +
           \half\deg\pi \deg{\bar\pi}
            \deg{\xi_l}\Big(\frac{1}{\rho^2}+\frac{1}{\rhb^2}\Big)\\
            &\pheq + \Big[\frac{\deg\pi}{\rho} (\edtht + \deg{\bar\alpha})
            +\frac{\deg{\bar\pi}}{\rhb} (\edthpt+\deg\alpha )\Big] \deg{\xi_l}
+
           \half\Big(\frac{1}{\rho}+\frac{1}{\rhb}\Big)\thpt\deg{\xi_l}\\
	    &\pheq -[\deg\tab \rho(\edtht+\deg{\bar\alpha} ) +
            \deg\tau \rhb(\edthpt+\deg\alpha )]\deg{\xi_l}
	     +\deg\tab \deg{\xi_m}\frac{\rho}{\rhb}
            +\deg\tau \deg{\xi_{\mb}} \frac{\rhb}{\rho}\\
            &\pheq-\deg\pi \deg{\xi_m} \frac{1}{\rhb^2}
            -\deg{\bar\pi} \deg{\xi_{\mb}} \frac{1}{\rho^2}
	    -\deg\alpha\deg{\xi_m}\frac{1}{\rhb}
           -\deg{\bar\alpha}\deg{\xi_{\mb}}\frac{1}{\rho},\\
	 \xi_m &= \deg{\xi_m}\frac{1}{\rhb} -
         \deg{\bar\pi}\deg{\xi_l}\frac{1}{\rho} +
         \deg\tau\deg{\xi_l}\rhb -
         (\edtht+\deg{\bar\alpha})\deg{\xi_l}\!,\\
	 \xi_{\mb} &= \deg{\xi_{\mb}}\frac{1}{\rho} -
         \deg\pi \deg{\xi_l} \frac{1}{\rhb} + \deg\tab\deg{\xi_l}\rho -
         (\edthpt+\deg\alpha)\deg{\xi_l}\!,}\label{eqn:GENSOLGAUGE}
\end{equation}
where the quantities $\deg\Psi$, $\deg\tau$, $\deg\pi$ and $\deg\alpha$
determine properties of the background spacetime.\footnote{For example,
$\deg\pi\neq 0$ leads to the accelerating C-metrics.  The condition $\deg\pi=0$
implies $\deg\alpha=0$ and so $\deg\alpha$ is also related to parameters in the
C-metric.}  Now the gauge transformation for $h_{m\mb}$ becomes
\begin{equation}\fl
\eqalign{ h_{m\mb} &= \frac{\rho}{\rhb}\Big[\edthpt\deg{\xi_m} +
\thpt\deg{\xi_l}-\deg{B}\Big] +
\frac{\rhb}{\rho}\Big[\edtht\deg{\xi_{\mb}}  +
\thpt\deg{\xi_l}+\deg{B}\Big]\\
&\pheq +
(\rho+\rhb) \Big[-\half(\edthpt\edtht+\edtht\edthpt-\deg{\rhp}-
\rhb^{\prime\circ})\deg{\xi_l}
+\deg{\xi_n}-\deg{A}\Big],}
\label{eqn:GENTYPEDSOL}
\end{equation}
where we have introduced (note that $\deg{B}$ is purely imaginary)
\begin{equation}
\eqalign{\deg{A} &= \half\{2\deg{\alpha}\edtht + \edtht(\deg{\alpha})
+ \deg{\alpha}\deg{\bar\alpha} - \deg{\pi}\deg{\tau}\}\deg{\xi_l} +
\hbox{c.c.},\\
\deg{B} &= \hbox{$1\over 4$}\{4\deg{\pi}\edtht + \edtht(\deg{\pi}) +
5\deg{\bar\alpha}\deg{\pi}-2\deg{\pi}\deg{\bar\pi}\deg{\Omega}\}\deg{\xi_l}
- \hbox{c.c.},}\label{eqn:cmetstuff}
\end{equation}
with c.c. indicating the complex conjugate.  Integration of the backgrounds
where $\deg\pi\neq 0$ and $\deg\alpha\neq 0$ using the Held technique has not
made its way into the literature and is beyond the scope of the present work.
As a result, derivatives of $\deg{\pi}$ and $\deg{\alpha}$ appear in
\eref{eqn:cmetstuff} but do no harm to our argument.  Choosing any gauge vector
that satisfies
\begin{equation}
\fl{ \edthpt\deg{\xi_m}+\thpt\deg{\xi_l}-\deg{B}=\deg{a},\qquad
	  -\half(\edthpt\edtht+\edtht\edthpt-\deg{\rhp}-\rhb^{\prime\circ}
	  )\deg{\xi_l} + \deg{\xi_n}
	  -\deg{A}
	  =\deg{b}, }\label{eqn:IRGTD}
\end{equation}
will serve to impose the trace condition in the full IRG.  Once again we have
established that ${\cal T}_{ll}=0$ is both a necessary and sufficient condition
for the existence of the full IRG.  Note that by setting $\deg\pi=\deg\alpha=0$
(i.e. ignoring the C-metrics) in the background, $\deg A=\deg B=0$, and the
result is virtually identical to \eref{eqn:GENTYPEIISOL} and \eref{gauge:IRG}.
There is one simplification in that now
$\rhb^{\prime\circ}=\deg{\rhp}$\cite{Stewart:1974}.
The full extent of the remaining residual gauge freedom in \eref{eqn:IRGTD}
can be demonstrated along the same lines as used in section \ref{typeII:RGF}.
As for the case of a type II background, it resides chiefly in the freely
specifiable $\deg{\xi_m}$ and $\deg{\xi_{\mb}}$.

\section{Imposing the IRG in Schwarzschild spacetime}\label{Schwarz:IRG}

We now demonstrate these results for Schwarzschild spacetime using
conventional, spherically symmetric coordinates, in which the
background metric takes the form:
\begin{equation}
ds^{2}=f(r)\,dt^{2}-dr^{2}\!/\!f(r)-r^{2}\left(d\theta^{2}+\sin^{2}\theta\,
d\phi^{2}\right),\label{1}\end{equation}
where we have introduced $f(r)\!\!=\!\!(1-2M/r)$.  Metric perturbations,
$h_{ab}(t,r,\theta,\phi)$, about the Schwarzschild geometry can be
expressed in terms of the RW decomposition\cite{Regge:1957td}.  We decompose
the angular dependence of these perturbations into spherical harmonics and the
time dependence into constant frequency Fourier modes:
\begin{equation}
h_{ab}(t,r,\theta,\phi)=\sum_{\ell,m,\omega} e^{-i\omega
t}\,h_{ab}^{\ell m\omega}(r){\cal O}_{ab}(\theta,\phi)Y_{\ell m}(\theta,\phi),
\end{equation}
where ${\cal O}_{ab}(\theta,\phi)$ is an angular operator (see
\cite{Thorne:1980} for details) and there is no sum here on the spacetime
indices $\{ab\}$.  With respect to rotation of the background coordinate
system, $h_{tt},\,\!h_{rr}\,{\rm and~} h_{tr}$ transform as scalars, $\{
h_{t\theta},\, h_{t\phi}\}$ and $\{ h_{r\theta},\, h_{r\phi}\}$ transform as a
pair of vectors on the 2-sphere and $\{
h_{\theta\theta},\, h_{\theta\phi},\, h_{\phi\phi}\}$ transforms as a symmetric
covariant tensor on the 2-sphere.

It is well known that the components of the metric perturbation decouple into
two classes, labeled as even and odd parity,  according to their behavior under
a parity transformation $P:$($\theta\rightarrow\pi-\theta$,
$\phi\rightarrow\pi+\phi$).  Out of the ten independent components of the
metric perturbation for each mode (of specific $\{\ell,m,\omega\}$), the even
parity perturbations, for which $P=(-1)^{\ell}$, have seven independent
components and are given by $h_{ab}^{\mathrm{even}}$,
\begin{equation*}
\fl e^{-i\omega t}\!\left[\begin{array}{cccc}
f(r)H_{0}(r) & H_{1}(r) & {}^eh_{0}(r){\partial\over \partial\theta} &
{}^eh_{0}(r)\frac{\partial}{\partial\phi}\\
\mathrm{sym} & H_{2}(r)/f(r) & {}^eh_{1}(r)\frac{\partial}{\partial\theta} &
{}^eh_{1}(r)\frac{\partial}{\partial\phi}\\
\mathrm{sym} & \mathrm{sym} &
r^{2}\left[K(r)+G(r)\frac{\partial^{2}}{\partial\theta^{2}}\right]
&r^{2}G(r)\left[\frac{\partial^{2}}{\partial\theta\partial\phi}-
\frac{\cos\theta}{\sin\theta}\frac{\partial}{\partial\phi}\right] \\\bs
\mathrm{sym} & \mathrm{sym} & \mathrm{sym} &
\matrix{r^{2}G(r)\left[\frac{\partial^{2}}{\partial\phi^{2}}+
\frac{\sin2\theta}{2}\frac{\partial}{\partial\theta}\right]\cr
 +r^{2}K(r)\sin^{2}\theta\cr}\end{array}\right]\!Y_{\ell m}(\Omega).
 \end{equation*}
The odd parity perturbations, with $P=(-1)^{l+1}$, have three independent
components, and are given by $h_{ab}^{\mathrm{odd}}$,
\begin{equation*}
\fl e^{-i\omega t}\!\left[\begin{array}{cccc}
0 & 0 & -{}^oh_{0}(r)\frac{1}{\sin\theta}\frac{\partial}{\partial\phi} &
{}^oh_{0}(r)\sin\theta\frac{\partial}{\partial\theta}\\
0 & 0 & -{}^oh_{1}(r)\frac{1}{\sin\theta}\frac{\partial}{\partial\phi} &
{}^oh_{1}(r)\sin\theta\frac{\partial}{\partial\theta}\\
\mathrm{sym} & \mathrm{sym} &
h_{2}(r)\left[\frac{1}{\sin\theta}
\frac{\partial^{2}}{\partial\theta\partial\phi}-
\frac{\cos\theta}{\sin^{2}\theta}\frac{\partial}{\partial\phi}\right] &
\matrix{\frac{1}{2}h_{2}(r)\left[\frac{1}{\sin\theta}
\frac{\partial^{2}}{\partial\phi^{2}}+\right.\cr\left.\cos\theta
\frac{\partial}{\partial\theta}-\sin\theta
\frac{\partial^{2}}{\partial\theta^{2}}\right]\cr}\\
\mathrm{sym} & \mathrm{sym} &\mathrm{sym}
&-h_{2}(r)\left[\sin\theta\frac{\partial^{2}}{\partial\theta\partial\phi}-
\cos\theta\frac{\partial}{\partial\phi}\right]\end{array}
\right]\!Y_{\ell m}(\Omega).
\end{equation*}

In the background spacetime, the Einstein tensor is identically zero. The
perturbed Einstein tensor (which includes contributions from the metric
perturbation $h_{ab}$ up to first order) transforms in the same way as the
metric perturbations with respect to rotations on the 2-sphere.  Hence, it has
the same angular decomposition as the metric perturbation. For the even parity,
we write the perturbed Einstein tensor, ${\cal E}_{ab}^{\mathrm{even}}$, as
\begin{equation*}
\fl e^{-i\omega t}\!\left[\begin{array}{cccc}
f(r)E_{1}(r) & E_{2}(r) &
E_{4}(r)\frac{\partial}{\partial\theta} &
E_{4}(r)\frac{\partial}{\partial\phi}\\
\mathrm{sym} & E_{3}/f(r) &
E_{5}(r)\frac{\partial}{\partial\theta} &
E_{5}(r)\frac{\partial}{\partial\phi}\\
\mathrm{sym} & \mathrm{sym} &
r^{2}\left[E_{6}(r)+E_{7}(r)\frac{\partial^{2}}{\partial\theta^{2}}\right]
&r^{2}E_{7}(r)\left[\frac{\partial^{2}}{\partial\theta\partial\phi}-
\frac{\cos\theta}{\sin\theta}\frac{\partial}{\partial\phi}\right]\\\bs
\mathrm{sym} & \mathrm{sym} & \mathrm{sym} &
\matrix{r^{2}E_{7}(r)\left[\frac{\partial^{2}}{\partial\phi^{2}}+
\frac{\sin2\theta}{2}\frac{\partial}{\partial\theta}\right]\cr
+r^{2}E_{6}(r)\sin^{2}\theta\cr}\end{array}\right]\!Y_{\ell m}(\Omega),
 \end{equation*}
and for odd parity we write perturbed Einstein tensor,
${\cal E}_{ab}^{\mathrm{odd}}$, as
\begin{equation*}
\fl e^{-i\omega t}\!\left[\begin{array}{cccc}
0 & 0 & -F_{1}(r)\frac{1}{\sin\theta}\frac{\partial}{\partial\phi} &
F_{1}(r)\sin\theta\frac{\partial}{\partial\theta}\\
0 & 0 & -F_{2}(r)\frac{1}{\sin\theta}\frac{\partial}{\partial\phi} &
F_{2}(r)\sin\theta\frac{\partial}{\partial\theta}\\
\mathrm{sym} & \mathrm{sym} &
F_{3}(r)\left[\frac{1}{\sin\theta}
\frac{\partial^{2}}{\partial\theta\partial\phi}-
\frac{\cos\theta}{\sin^{2}\theta}\frac{\partial}{\partial\phi}\right]
&\matrix{\frac{1}{2}F_{3}(r)\left[\frac{1}{\sin\theta}
\frac{\partial^{2}}{\partial\phi^{2}}+\right.\cr\left.\cos\theta
\frac{\partial}{\partial\theta}-\sin\theta
\frac{\partial^{2}}{\partial\theta^{2}}\right]\cr}\\
\mathrm{sym} & \mathrm{sym} & \mathrm{sym} &
-F_{3}(r)\left[\sin\theta\frac{\partial^{2}}{\partial\theta\partial\phi}-
\cos\theta\frac{\partial}{\partial\phi}\right]\end{array}
\right]\!Y_{\ell m}(\Omega).
\end{equation*}

The spherical symmetry and the static nature of the background geometry ensures
that the perturbed Einstein equations decouple into individual modes of
$\{\ell,m,\omega,P\}$. That is, each component of the perturbed ${\cal E}_{ab}$
belonging to a specific $\{\ell,m,\omega,P\}$ mode depends only on the metric
perturbations of the same mode, $h_{ab}^{\ell m\omega P}$. Hence, it is
generally sufficient to consider a single mode of the metric perturbation for
our analysis.

We now impose the $l\!\cdot\!h$ gauge conditions \eref{eqn:IRG} on a specific
mode of the perturbed metric.  We have, $l^{a}h_{ab}=0$, where
$l^{a}=\left(1/f(r),1,0,0\right)$ from equations \eref{eqn:tetrad} below, and
$l^a$ is a repeated PND of the Schwarzschild background.  For the odd parity
perturbations, we can write $h_{0}$ in terms of $h_{1}$.
\begin{equation}
{}^oh_{0}(r)=-f(r)\,{}^oh_{1}(r).\label{eqn:oddIRG}
\end{equation}
For the even parity perturbations,
\begin{equation}
H_{0}(r)=-H_{1}(r)=H_{2}(r),\quad{\rm and}\quad
{}^eh_{0}(r)=-f(r)\,{}^eh_{1}(r).\label{eqn:evenIRG}
\end{equation}
The trace of the metric perturbations is a scalar with respect to rotation on a
sphere of constant $r$ and constant $t$. Hence, it can be written as
\begin{equation}
h_{ab}g^{ab}=\sum_{\ell,m,\omega,P} e^{-i\omega t}\,T^{\ell m\omega P}(r)\,
Y_{\ell m}(\theta,\phi),
\end{equation}
Expanding the LHS gives,
\begin{equation}
h_{ab}g^{ab}=\frac{1}{f(r)}h_{tt}-f(r)h_{rr}-\frac{1}{r^{2}}h_{\theta\theta}
-\frac{1}{r^{2}\sin^{2}\theta}h_{\phi\phi}.
\end{equation}
Note that the trace vanishes for the odd parity perturbations, while for the
even parity the trace is equal to (we suppress labels \{$\ell,m,\omega,P$\}
when
ambiguity is unlikely)
\begin{equation}
T(r)=-\left[2K(r)-\ell(\ell+1)G(r)\right].
\end{equation}
We have used the fact that the spherical harmonics are eigenfunctions of the
angular momentum operator:
\begin{equation}
\left[\frac{\partial^{2}}{\partial\theta^{2}}+
\frac{1}{\sin^{2}\theta}\frac{\partial^{2}}{\partial\phi^{2}}+
\frac{\cos\theta}{\sin\theta}\frac{\partial}{\partial\theta}\right]
Y_{\ell m}(\theta,\phi)=-\ell(\ell+1)Y_{\ell m}(\theta,\phi).\label{eqn:AM}
\end{equation}
The perturbed Einstein tensor obtained in this gauge is given in \ref{Schwarz}.

\subsection{Residual gauge freedom in the Schwarzschild geometry}

To determine the residual gauge freedom in the Schwarzschild background we
require, as in (\ref{eqn:MPGT}),
\begin{equation}
l^{a}h_{ab}=l^{a}\xi_{(a;b)}.
\end{equation}
Writing $l^{a}h_{ab}$ as $B_{b}$ and using the metric (\ref{1}) to compute the
covariant derivatives, gives
\begin{eqnarray}
\eqalign{B_{t}= &
[2\xi_{t,t}-f'(r)\xi_{t}]/f(r)+\xi_{r,t}+\xi_{t,r}-f'(r)\xi_{r},\\
B_{r}= & [\xi_{t,r}+\xi_{r,t}+f'(r)\{\xi_{r}-\xi_{t}/f(r)\}]/f(r)+2\xi_{r,r},\\
B_{\theta}=
&[\xi_{t,\theta}+\xi_{\theta,t}]/f(r)+\xi_{r,\theta}+
\xi_{\theta,r}-{2}\xi_{\theta}/r,\\
B_{\phi}= &
[\xi_{t,\phi}+\xi_{\phi,t}]/f(r)+\xi_{r,\phi}+\xi_{\phi,r}-{2}\xi_{\phi}/r.}
\label{g1}
\end{eqnarray}
Gauge vectors, $\xi_{a}$, which correspond to residual gauge freedom, solve the
above equations with $B_{a}=0$.  Moreover, since we are going to deal with the
metric perturbations of one single mode (specific $\{\ell,m,\omega,P\}$) at a
time, we want $h_{ab}$ and $\xi_a$ to correspond to the same mode.  This
restricts the functional form of our gauge vector $\xi_{a}(x^{b})$.

For even parity perturbations, we consider an even gauge vector of the form,
\begin{eqnarray}
\eqalign{
\xi_{t}=&-e^{-i\omega t}\,P(r)Y_{\ell m}(\theta,\phi),\\
\xi_{r}=&e^{-i\omega t}\,R(r)Y_{\ell m}(\theta,\phi),\\
\xi_{\theta}=&e^{-i\omega t}\,rS(r)\frac{\partial}{\partial\theta}Y_{\ell
m}(\theta,\phi),\\
\xi_{\phi}=&e^{-i\omega t}\,rS(r)\frac{\partial}{\partial\phi}Y_{\ell
m}(\theta,\phi).}\label{even:gauge}
\end{eqnarray}
For odd parity perturbations, we consider an odd gauge vector of the form,
\begin{eqnarray}
\eqalign{
\xi_{t}=&0,\\
\xi_{r}=&0,\\
\xi_{\theta}=&-e^{-i\omega
t}\,Q(r)\frac{1}{\sin\theta}\frac{\partial}{\partial\phi}Y_{\ell
m}(\theta,\phi),\\
\xi_{\phi}=&e^{-i\omega
t}\,Q(r)\sin\theta\frac{\partial}{\partial\theta}Y_{\ell
m}(\theta,\phi).}\label{odd:gauge}
\end{eqnarray}
Inserting these forms of gauge vector into (\ref{g1}) and taking $B_{a}=0$, we
arrive at equations for the residual gauge freedom.  For the even parity gauge
vector, we have
\begin{eqnarray}
\eqalign{
\fl0= & e^{-i\omega t}\left[\{{2i\omega}P(r)+f'(r)P(r)\}/f(r)-i\omega
R(r)-P'(r)-f'(r)R(r)\right]Y_{\ell m}(\theta,\phi),\\
\fl0= & e^{-i\omega t}\left[\{f'(r)[R(r)+P(r)/f(r)]-i\omega
R(r)-P'(r)\}/f(r)+2R'(r)\right]Y_{\ell m}(\theta,\phi),\\
\fl0= & e^{-i\omega t}\left[-\{i\omega
rS(r)+P(r)\}/f(r)+R(r)-S(r)+rS'(r)\right]\frac{\partial}{\partial\theta}Y_{\ell
m}(\theta,\phi),\\
\fl0= &
e^{-i\omega t}\left[-\{i\omega
rS(r)+P(r)\}/f(r)+R(r)-S(r)+rS'(r)\right]\frac{\partial}{\partial\phi}Y_{\ell
m}(\theta,\phi).}
\end{eqnarray}
For the odd parity gauge vector, we have
\begin{eqnarray}
\eqalign{
0=e^{-i\omega t}\left[i\omega
Q(r)/f(r)-Q'(r)+2Q(r)/r\right]\frac{1}{\sin\theta}\frac{\partial}{\partial\phi}
Y_{\ell m}(\theta,\phi),\\
0=-e^{-i\omega t}\left[i\omega
Q(r)/f(r)-Q'(r)+2Q(r)/r\right]\sin\theta\frac{\partial}{\partial\theta}Y_{\ell
m}(\theta,\phi).}
\end{eqnarray}

These equations can be solved completely for the functions $P(r)$,
$R(r)$, $S(r)$ and  $Q(r)$ in terms of four arbitrary constants:
\begin{eqnarray}
\eqalign{
P(r)= e^{i\omega r_{*}}\left[C_{1}-C_{2}\left(i\omega r+f(r)\right)\right],\\
R(r)= e^{i\omega r_{*}}\left[C_{1}-i\omega rC_{2}\right]/f(r),\\
S(r)= e^{i\omega r_{*}}\left[C_{2}+C_{3}r\right],{\rm and}\\
Q(r)=e^{i\omega r_{*}}D\,r^{2},}\label{odd:rgf}
\end{eqnarray}
where we have
introduced
$r_{*}=r+2M\ln(r/2M-1)$.  These solutions correspond to the residual gauge
freedom of the light-cone
gauge in \cite{Preston:2006ze}.

\subsection{Condition for the Trace to vanish}\label{Strace}

The residual gauge freedom can be used
to change the trace of the metric perturbation by a quantity
$\mathrm{Tr}(\xi_{(a;b)})=g^{ab}\xi_{(a;b)}$. For an odd parity
perturbation, this quantity is easily seen to be zero. For an even parity
perturbation, this quantity is evaluated to be
\begin{equation}
\fl \mathrm{Tr}(\xi_{(a;b)})=e^{-i\omega(t-r_{*})}\!\left(2i\omega
C_{2}+\ell(\ell+1)C_{3}-\frac{2C_{1}-\ell(\ell+1)C_{2}}{r}\right)\!Y_{\ell
m}(\theta,\phi).\label{SgaugeT}
\end{equation}
A particular linear combination, $E_{1}+2E_{2}+E_{3}$, of the Einstein tensor
components given in \ref{Schwarz} is exactly $f(r){\cal E}_{ll}$, and reduces
to a second order operator acting only on the variable $T(r)$.  For
perturbations satisfying ${\cal T}_{ll}=0$, the $ll$-component of
\eref{eqn:pEEs} gives:
\begin{equation}
\fl2{\cal E}_{ll}=-T''(r)+\frac{2(i\omega
r^{2}-r+2M)}{r(r-2M)}T'(r)+\frac{2i\omega (r-3M)+\omega^2
r^{2}}{(r-2M)^{2}}T(r)=0.
\end{equation}
The solution to this equation is obtained in terms of two arbitrary constants
$A,B$:
\begin{equation}
T(r)=e^{i\omega r_{*}}(A+{B}/{r}).\label{StraceT}
\end{equation}

{}From \eref{SgaugeT}, we already know the degrees of freedom that exist in the
trace of the metric perturbation due to residual gauge freedom. One sees that
the arbitrary constants $C_{1},\, C_{2},\, C_{3}$ can be chosen to exactly
cancel
$A$ and $B$.  This analysis confirms that the residual gauge freedom can
generally be exploited to set the trace of the metric perturbations to zero for
perturbations with ${\cal T}_{ll}=0$, and verifies that ${\cal E}_{ll}=0$ is a
sufficient condition for constructing an IRG.  Once this has been done, there
still exists one constant residual degree of gauge freedom per mode of metric
perturbation (both even and odd), just as in sections \ref{typeII} and
\ref{typeD}. It is not clear how to fix these degrees of freedom in order to
get some more useful analytical property of the metric perturbations.

Note that, in \eref{SgaugeT}, when $\ell=0$, two of the four terms vanish
identically.  Next note that, for $\omega=0$, another term also vanishes.  A
time independent perturbation for which both $\ell$ and $\omega$ are zero,
corresponds to a change in mass of the black hole.  From \eref{SgaugeT}, it
would then appear that, for such a static mass perturbation, we can no longer
cancel the A-term from the trace in \eref{StraceT}.  This is a mere artifact of
our analysis and is not a fundamental obstacle.  Use of the Fourier transform
in $t$ is not permitted unless the perturbations belong in some suitable
function space, say $L^2$. Polynomials would fail this test, but they are
required in the gauge transformation for this case\cite{Preston:2006ze}.
Allowing polynomials in $t$ and restoring the time derivative in \eref{SgaugeT}
corrects this defect.

\subsection{Connection with the GHP Formulation}
Now that we have performed the same analysis for both a generic type D and the
Schwarzschild backgrounds, we are in a position to make a direct comparison.
For that purpose, we introduce a complete set of null tetrad vectors:
\begin{equation}
\fl l^{a}=(1/f(r),1,0,0),\quad n^{a}=\frac{1}{2}(1,-f(r),0,0),\quad
 m^{a}=\frac{1}{\sqrt{2}r}(0,0,1,i/\sin\theta).\label{eqn:tetrad}
\end{equation}
With this choice (introduced by Kinnersley\cite{kinnersley:1195}), the tetrad
components of the residual gauge vector in the Schwarzschild spacetime become,
after inserting \eref{odd:rgf} into \eref{even:gauge} and \eref{odd:gauge}:
\begin{equation}
\eqalign{
\xi_{l}=\xi_{a}l^{a}=e^{-i\omega(t-r_{*})}C_{2}Y_{\ell m}(\theta,\phi),\\
\xi_{n}=\xi_{a}n^{a}=
-e^{-i\omega(t-r_{*})}[C_{1}-C_{2}(ir\omega+f(r)/2)]Y_{\ell m}(\theta,\phi),\\
\xi_{m}=\xi_{a}m^{a}=
\frac{1}{\sqrt{2}}e^{-i\omega(t-r_{*})}[C_{2}+rC_{3}+irD]\!
\left[\frac{\partial}{\partial\theta}
+ \frac{i}{\sin\theta}\frac{\partial}{\partial\phi}\right]\!Y_{\ell
m}(\theta,\phi).}\label{eqn:schwgvtet}
\end{equation}
Before carrying out the comparison, we need to express the null derivative
operators in the Kinnersley tetrad \eref{eqn:tetrad}.  Acting on a scalar
quantity, the operator $\th$ is given by
\begin{equation}
\th = l^a\Theta_a = \frac{1}{f(r)}\frac{\partial}{\partial t} +
\frac{\partial}{\partial r}.
\end{equation}
Therefore, in the Schwarzschild geometry, quantities annihilated by $\th$ have
the form
\begin{equation}
\deg x = \deg{x}(t-r_*,\theta,\phi).
\end{equation}
Also, in the Schwarzschild geometry, $\rhb=\rho=-1/r$ and $\tau=\tap=0$.  So,
when acting on scalars of spin-weight $s$, the operators $\edtht$ and $\edthpt$
are (see \eref{eqn:O2} and \eref{eqn:O3} and also \cite{Goldberg:1966uu})
\begin{equation}
\eqalign{
\edtht &= \frac{\edth}{\rhb} = -rm^a\Theta_a =
-\frac{(\sin\theta)^{s}}{\sqrt{2}}\Big(\frac{\partial}{\partial\theta} +
\frac{i}{\sin\theta}\frac{\partial}{\partial\phi}\Big)(\sin\theta)^{-s},\\
\edthpt &= \frac{\edthp}{\rho} = -r\mb^a\Theta_a =
-\frac{(\sin\theta)^{-s}}{\sqrt{2}}\Big(\frac{\partial}{\partial\theta} -
\frac{i}{\sin\theta}\frac{\partial}{\partial\phi}\Big)(\sin\theta)^{s}.}
\label{eqn:edths}
\end{equation}
Since ${\bar\Psi}_2=\Psi_2=\deg\Psi\rho^3=-M/r^3$, we similarly have, for
$\thpt$ acting on scalars of boost-weight $b$ (see
\eref{eqn:O1} in \ref{Held}, and \cite{GHP}):
\begin{equation}
\eqalign{
\thpt &= \thp + \frac{(p+q)\Psi_2}{2\rho} = n^a\Theta_a
+b\deg\Psi\rho^2 \\\bs
&=\frac{f(r)^{-b}}{2}\Big(\frac{\partial}{\partial t} -
f(r)\frac{\partial}{\partial r} +
b\frac{2M}{r^2}\Big)f(r)^{b}=\frac{1}{2}\Big(\frac{\partial}{\partial t} -
f(r)\frac{\partial}{\partial r}\Big).}\label{eqn:thptc}
\end{equation}
Now, given that, in the Schwarzschild background,
$\deg\tau=\deg\pi=\deg\alpha=0$ and $\deg\Psi=M$ we can write
\eref{eqn:GENSOLGAUGE} as
\begin{equation}
\eqalign{
\xi_l &= \deg{\xi}_l,\\
\xi_n &= \deg{\xi}_n - \frac{r}{2}\Big(\frac{\partial}{\partial t} -
f(r)\frac{\partial}{\partial r}+\frac{2M}{r^2}\Big)\deg{\xi}_l,\\
\xi_m &= -r\deg{\xi}_m +
\frac{1}{\sqrt{2}}\Big(\frac{\partial}{\partial\theta} +
\frac{i}{\sin\theta}\frac{\partial}{\partial\phi}\Big)\deg{\xi}_l.}
\label{eqn:gaugeschw}
\end{equation}
Comparing \eref{eqn:schwgvtet} and \eref{eqn:gaugeschw}, it is clear
that\footnote{To obtain these relations, we have used $rf'(r)=1-f(r)$, which
holds in the Schwarzschild spacetime.}
\begin{equation}
\eqalign{
\deg{\xi}_l &= e^{-i\omega(t-r_{*})}C_{2}Y_{\ell m}(\theta,\phi),\\
\deg{\xi}_n &= -e^{-i\omega(t-r_{*})}(C_{1}-C_{2}/2)Y_{\ell m}(\theta,\phi),\\
\deg{\xi}_m &= -e^{-i\omega(t-r_{*})}\frac{C_{3}+iD}{\sqrt{2}}
\Big(\frac{\partial}{\partial\theta}
+ \frac{i}{\sin\theta}\frac{\partial}{\partial\phi}\Big)Y_{\ell
m}(\theta,\phi).}\label{eqn:ghpcorr}
\end{equation}
To conclude, by comparing \eref{StraceT} and \eref{eqn:hmmbsoln}, and recalling
$\rhb=\rho=-1/r$, we see that with
\begin{equation}
{A=\deg a +\deg {\bar a}\!,\quad{\rm and}\quad B=-2\deg b\!,}
\end{equation}
the equivalence of the two formulations is established.  Finally, with $\deg
\rhp=-\half$, and using \eref{eqn:rhp}, \eref{eqn:ghpcorr} and \eref{eqn:AM} in
\eref{eqn:TCOMP}, we can demonstrate its complete correspondence with
\eref{SgaugeT}.

\section{Discussion}
We have concentrated on Petrov type II spacetimes in this paper because they
satisfy a minimum requirement necessary for the existence of a radiation gauge,
namely the occurrence of a repeated PND.  With our new form of the perturbed
Einstein equations, use of NP methods has allowed us to treat this general
class of spacetimes without either choosing coordinates or finding a metric.
In this context, the Held technique has allowed us to exploit our form of the
equations by enabling partial integration in solving ${\cal E}_{ll}=0$ while
investigating the existence of the IRG.  Additionally, the Held technique has
allowed us to completely characterize the residual gauge freedom and use it in
the radiation gauge construction.  By explicit demonstration, this work
establishes our new form of the perturbed Einstein equations as a powerful tool
within perturbation theory, both in conjunction with the Held technique and
otherwise.

For perturbations with ${\cal T}_{ll}=0$, our characterization of the residual
gauge freedom is sufficiently complete that we can explicitly demonstrate the
required gauge choice to remove any non-zero solution for the trace obtained
via ${\cal E}_{ll}=0$.  Thus, in type II spacetimes, radiation gauges can be
established by a genuine gauge choice, even if only after a solution of ${\cal
E}_{ll}=0$ is chosen.

There are subtle differences between the general type II case and the more
restricted type D case, as there are also in the construction of Hertz
potentials for the two cases.  Stewart\cite{Stewart} writes out the type II
case rather fully for an IRG.  In this case, the perturbation in $\Psi_0$ is
tetrad and gauge invariant, while the potential satisfies the adjoint (in the
sense detailed by Wald\cite{Wald:1978vm}) of the $s=+2$ Teukolsky equation.
Remarkably, in the type D case, this adjoint is actually the $s=-2$ Teukolsky
equation, also satisfied by the gauge and tetrad invariant perturbation in
$\Psi_4$.  In the type II case, the adjoint equation is the same as in type D,
but $\Psi_4$ is no longer tetrad invariant.  Compared to the type D result, the
expression for $\Psi_4$ given by Stewart has many extra terms depending on
$\kappa'$ and $\sigma'$, so presumably it does not satisfy the same equation as
the potential.  As a consequence, metric reconstruction would be restricted to
being built around the perturbation for $\Psi_0$.

In the context of a specific type D background, Wald\cite{Wald:73} has argued
that mass and angular momentum perturbations are not given by any solution to
the Teukolsky equations, and Stewart\cite{Stewart} has shown that these cannot
be represented in a radiation gauge in terms of a potential.  What we have done
is identify the gauge freedom which remains in the fully satisfied radiation
gauges, neither interfering with the radiation gauge prescription nor ruling
out the possibility of mass and angular momentum perturbations.  By realizing
the explicit construction of the radiation gauges for type II background
spacetimes and by identifying the remaining gauge freedom which they allow, we
have, in a sense, completed a task initially embarked upon by
Stewart\cite{Stewart}, though in the different context of Hertz potentials.

\ack
This work has been supported in part by NSF Grants PHY-0245024  and
PHY-0555484 with the University of Florida.  L.R.P.~acknowledges support
from the Alumni Fellowship Program at the University of Florida during the
early phases of the work reported here.  K.S.~acknowledges support from an IFT
Summer Fellowship at the University of Florida during part of this work.

\appendix
\section{The Perturbed Einstein Equations for a type II
Background}\label{App:typeII}

In this appendix we write the components of the perturbed Einstein tensor for
an arbitrary type II background.  We have assumed the PND is aligned with $l^a$
and made use of the Goldberg-Sachs theorem.  Note that the equations for ${\cal
E}_{lm}$, ${\cal E}_{nm}$ and ${\cal E}_{mm}$ are complex, so ${\cal
E}_{l\mb}={\bar{\cal E}}_{lm}$ and so on:

\begin{equation}
\fl\eqalign{
{\cal E}_{ll} &= \{(\edthp-\tap)(\edth-\tab') + \rho(\thp+\rhp-\rhb') -
(\th-\rho)\rhp + \Psi_2\}h_{ll}\\
&\pheq + \{-(\rho+\rhb)(\th+\rho+\rhb) +4\rho\rhb\}h_{ln}\\
&\pheq + \{-(\th-3\rhb)(\edthp-\tap+\tab) + \tab\th-\rhb\edthp\}h_{lm}\\
&\pheq + \{-(\th-3\rho)(\edth+\tau-\tab') + \tau\th-\rho\edth\}h_{l\mb}\\
&\pheq + \{\th(\th-\rho-\rhb)+2\rho\rhb\}h_{m\mb},}\label{eqn:Ell_app}
\end{equation}

\begin{equation}
\fl\eqalign{
{\cal E}_{nn} &= \{2\kap\kab'\}h_{ll}\\
&\pheq + \{(\edthp-\tab)(\edth-\tau) +\rhb'(\th-\rho+\rhb) -
(\thp-\rhb')\rhb + {\bar\Psi}_2 +2\rhb\rhb'\}h_{nn}\\
&\pheq + \{-(\rhp+\rhb')(\thp+\rhp+\rhb') + 4\rhp\rhb'
 - (\edthp-2\tab)\kab' -(\edth-2\tau)\kap\}h_{ln}\\
&\pheq + \{(\thp-\rhb')\kap +\kap(\thp-\rhp-\rhb')
-\kab'\sip\}h_{lm}\\
&\pheq + \{(\thp-\rhp)\kab' +\kab'(\thp-\rhb'-\rhp) -\kap\sib'\}h_{l\mb}\\
&\pheq + \{-(\thp-3\rhp)(\edthp+\tap-\tab) + \tap\thp - \rhp\edthp -\kap\th\\
&\pheq\pheq +(\th-2\rho+\rhb)\kap + (\edth-3\tau+\tab')\sip +\edth(\sip)
-\Psi_3\}h_{nm}\\
&\pheq + \{-(\thp-3\rhb')(\edth+\tab'-\tau) + \tab'\thp - \rhb'\edth
-\kab'\th\\
&\pheq\pheq +(\th-2\rhb+\rho)\kab' + (\edthp-3\tab+\tap)\sib' +\edthp(\sib') -
{\bar \Psi_3}\}h_{n\mb}\\
&\pheq + \{-(\edthp-2\tab)\kap - \sip(\thp-\rhp+\rhb')\}h_{mm} \\
&\pheq + \{-(\edth-2\tau)\kab' - \sib'(\thp-\rhb'+\rhp)\}h_{\mb\mb}\\
&\pheq + \{\thp(\thp-\rhp-\rhb') + \kap(\tau-\tab') +
\kab'(\tab-\tap) +2\sip\sib' + 2\rhp\rhb'\}h_{m\mb},}
\end{equation}

\begin{equation}
\fl\eqalign{
{\cal E}_{ln} &= \half\{\rhp(\thp-\rhp) + \rhb'(\thp-\rhb') +
(\edth-2\tab')\kap + (\edthp-2\tap)\kab' + 2\sip\sib'\}h_{ll}\\
&\pheq + \half\{\rho(\th-\rho) + \rhb(\th-\rhb)\}h_{nn}\\
&\pheq + \half\{-(\edthp+\tap+\tab)(\edth-\tau-\tab') -
(\edthp\edth+3\tau\tap+3\tab\tab') + 2(\tab+\tap)\edth\\
&\pheq\pheq+ (\th-2\rhb)\rhp +(\thp-2\rhp)\rhb - \rhb'(\th+\rho) -
\rho(\thp+\rhb') -\Psi_2 -{\bar\Psi}_2\}h_{ln}\\
&\pheq + \half\{(\thp-2\rhb')(\edthp-\tap) + \tab(\thp+\rhp+\rhb')
-\tap(\thp-\rhp)\\
&\pheq \pheq-(2\edthp-\tab)\rhb' - (\th-2\rhb)\kap +
\sip(\tau-\tab')\}h_{lm}\\
&\pheq + \half\{(\thp-2\rhp)(\edth-\tab') + \tau(\thp+\rhb'+\rhp)
-\tab'(\thp-\rhb')\\
&\pheq \pheq-(2\edth-\tau)\rhp - (\th-2\rho)\kab' +
\sib'(\tab-\tap)\}h_{l\mb}\\
&\pheq + \half\{(\th-2\rho)(\edthp-\tab) + (\tap+\tab)(\th+\rhb)
-2(\edthp-\tap)\rho-2\tab\th\}h_{nm}\\
&\pheq + \half\{(\th-2\rhb)(\edth-\tau) + (\tab'+\tau)(\th+\rho)
-2(\edth-\tab')\rhb-2\tau\th\}h_{n\mb}\\
&\pheq + \half\{-(\edthp-\tab)(\edthp-\tap) +
\tab(\tab-\tap)-\sip\rho\}h_{mm}\\
&\pheq + \half\{-(\edth-\tau)(\edth-\tab') +
\tau(\tau-\tab')-\sib'\rhb\}h_{\mb\mb}\\
&\pheq + \half\{(\edthp+\tap-\tab)(\edth-\tau+\tab') +
(\edthp\edth-\tau\tap-\tab\tab'+\tau\tab) - (\Psi_2+{\bar\Psi}_2)\\
&\pheq \pheq+(\thp-2\rhp)\rhb + (\th-2\rhb)\rhp +\rho(3\thp-2\rhb')
+\rhb'(3\th-2\rho)\\
&\pheq\pheq -2\thp\th + 2\rho\rhb' +2\edthp(\tau)-\tau\tab\}h_{m\mb},}
\end{equation}

\begin{equation}
\fl\eqalign{
{\cal E}_{lm} &= \half\{(\thp-\rhp)(\edth-\tab')
+(\edth-\tau-2\tab')\rhb' -(\edth-\tau)\rhp +\tau(\thp+\rhp)\\
&\pheq \pheq+\sib'(\edthp-\tap+\tab) + {\bar\Psi}_3+\rhb\kab'\}h_{ll}\\
&\pheq +\half\{-(\th-\rho+\rhb)(\edth+\tau-\tab') -
(\edth-3\tau+\tab')\rhb - 2\rho\tab'\}h_{ln}\\
&\pheq +\half\{-(\thp+\rhb')(\th-2\rhb) + \rho(\thp+2\rhp-2\rhb') -
4\rhp\rhb +2\Psi_2\\
&\pheq\pheq + (\edthp+\tab)(\edth-2\tab') - \tau(\edthp+\tap-2\tab)
-\tap(\tau-4\tab')\}h_{lm}\\
&\pheq + \half\{-\edth(\edth-2\tau) - \sib'(\th+2\rhb-4\rho) -
2\tab'(\tau-\tab')\}h_{l\mb}\\
&\pheq + \half\{\th(\th-2\rho) + 2\rhb(\rho-\rhb)\}h_{nm}\\
&\pheq + \half\{-(\th-\rhb)(\edthp-\tap+\tab) + 2\tab\rhb\}h_{mm}\\
&\pheq + \half\{(\th+\rho-\rhb)(\edth+\tab'-\tau) + 2\tab'(\th-2\rho) -
(\edth-\tau-\tab')\rhb +2\rho\tau\}h_{m\mb},}
\end{equation}

\begin{equation}
\fl\eqalign{
{\cal E}_{n\mb} &= \half\{(\thp-\rhp)\kap + \kap\thp +
\kab'\sip\}h_{ll}\\
&\pheq + \half\{(\th-\rho+\rhb)(\edthp-\tab) - (\edthp-2\tap+\tab)\rho +
\tap(\th-\rhb)\}h_{nn}\\
&\pheq + \half\{(-(\thp-\rhp+\rhb')(\edthp+\tap-\tab) -
(\edthp-3\tap+\tab)\rhb' + (\edth-\tau+\tab')\sip\\
&\pheq \pheq-2\sip\edth - \Psi_3-2\rhp\tab\}h_{ln}\\
&\pheq +\{\sip(\rhp-2\rhb')-\kap(\tap-2\tab)+\half\Psi_4\}h_{lm}\\
&\pheq +\half\{(\thp(\thp-2\rhp) + -\kap(\edth-2\tau+2\tab') +
\kab'(\edthp-4\tap+2\tab)\\
&\pheq \pheq+2\rhb'(\rhp-\rhb')+2\sip\sib'\}h_{l\mb}\\
&\pheq +\half\{-\edthp(\edthp-2\tap)+\sip(\th-2\rho+2\rhb) -
2\tab(\tap-\tab)\}h_{nm}\\
&\pheq +\half\{-(\thp-\rhb')(\th+2\rhb) + \rho(\thp-2\rhb') +
2\rhb'(\th-\rho) - \Psi_2 - 2{\bar\Psi}_2\\
&\pheq\pheq + (\edthp-3\tab)\edth + \tab'(2\edthp-\tab+4\tap)
-\tau(\edthp-2\tab)\}h_{n\mb}\\
&\pheq +\half\{-(\edthp-\tap)\sip - \sip\edthp\}h_{mm}\\
&\pheq +\half\{-(\thp-\rhb')(\edth-\tau+\tab') +2\tab'\rhb'
-\kab'(\th-2\rho+2\rhb) +\edthp(\sib')-\tab\sib'\}h_{\mb\mb}\\
&\pheq +\half\{(\thp+\rhp-\rhb')(\edthp-\tap+\tab) + 2\tab(\thp-2\rhp)
-(\edthp-\tap-\tab)\rhb'+2\rhp\tap \\
&\pheq\pheq+ (\edth-\tau-\tab')\sip + \sip\edth -\kap\th
-\Psi_3\}h_{m\mb},}
\end{equation}

\begin{equation}
\fl\eqalign{
{\cal E}_{mm} &= \{(\thp-2\rhp)\sib' +
\kab'(\edth+\tau-\tab')\}h_{ll}\\
&\pheq +\{-\edth(\edth-\tau-\tab') -2\tau\tab' +
\sib'(\rho-\rhb)\}h_{ln}\\
&\pheq +\{(\thp-\rhp)(\edth-\tab') - (\edth-\tau-\tab')\rhp +
\tau(\thp+\rhp-\rhb') -(\th-2\rhb)\kab'\\
&\pheq \pheq-\tab'(\th+\rhb')+\tab\sib' -{\bar\Psi}_3\}h_{lm}\\
&\pheq +\{-(\edth-\tau-\tab')\sib'-\sib'(\edth-\tau)\}h_{l\mb}\\
&\pheq +\{(\th-\rhb)(\edth-\tau) - (\edth-\tau-\tab')\rhb -\tau(\th+\rho)
+ \tab'(\th-\rho+\rhb)\}h_{nm}\\
&\pheq +\{-(\thp-\rhp)(\th-\rhb) + (\edth-\tau)\tap -
\tau(\edthp+\tap-\tab) + \Psi_2\}h_{mm}\\
&\pheq +\{(\th-2\rhb)\sib'+ (\tau+\tab')\edth +
(\tau-\tab')^2\}h_{m\mb},}
\end{equation}

\begin{equation}
\fl\eqalign{
{\cal E}_{m\mb} &= \half\{\thp(\thp-\rhp-\rhb') + 2\rhp\rhb' +
\kap(\tau-\tab') - \kab'(\tab-\tap) + 2\sip\sib'\}h_{ll}\\
&\pheq +\half\{\th(\th-\rho-\rhb) + 2\rho\rhb\}h_{nn}\\
&\pheq +\half\{-(\thp+\rhp-\rhb')(\th-\rho+\rhb) -\thp(\th+\rho)
+\rho(\thp+\rhp-\rhb') -{\bar\Psi}_2\\
&\pheq\pheq+(\edthp-\tab)(\edth-\tau-\tab') + \edthp\edth -
(\edth-2\tab')\tap -
\tab(2\edth+\tab')\\
&\pheq\pheq-2\tau(\edthp-\tab)+2\tap\tab'+\rhb\rhb'\}h_{ln}\\
&\pheq +\half\{-(\thp-2\rhp)(\edthp-2\tab) + \tab(\thp+2\rhp-2\rhb') +
2(\edth-\tab')\sip - \sip\edth\\
&\pheq \pheq-2\tap\rhb'-2\kap(\rho-\rhb)-\Psi_3\}h_{lm}\\
&\pheq +\half\{-(\thp-2\rhb')(\edth-2\tau) + \tau(\thp+2\rhb'-2\rhb) +
2(\edthp-\tau')\sib' - \sib'\edthp\\
&\pheq \pheq-2\tab'\rhp-2\kab'(\rhb-\rho)-{\bar\Psi}_3\}h_{l\mb}\\
&\pheq +\half\{-(\th-2\rhb)(\edthp-2\tap) + \tap(\th-2\rho-2\rhb) -
2\rho\tab+4\tap\rhb\}h_{nm}\\
&\pheq +\half\{-(\th-2\rho)(\edth-2\tab') + \tab'(\th-2\rhb-2\rho) -
2\rhb\tau+4\tab'\rho\}h_{n\mb}\\
&\pheq
+\half\{-\tab(\edthp-\tab)-\tap(\edthp-\tap)-(\th-2\rhb)\sip
\}h_{mm}\\
&\pheq +\half\{-\tau(\edth-\tau)-\tab'(\edth-\tab')-(\th-2\rho)\sib'
\}h_{\mb\mb}\\
&\pheq + \half\{2\thp\th -(\thp-\rhb')\rhb - (\th-\rho)\rhp
-\rho(\thp-\rhp+\rhb') - \rhb'(\th+\rho-\rhb)\\
&\pheq\pheq -(\edthp-2\tap)\tab' + \tau(\edthp+2\tab) -
\tap(\edth-\tab') + \tab(\edth+\tau) -\edthp(\tau)\\
&\pheq\pheq -\Psi_2-{\bar\Psi}_2\}h_{m\mb}.}
\end{equation}

\section{Integration \`a la Held{\rm
\cite{held:1974,held:1975,Stewart:1974}}}\label{Held}

We provide details of the integration that lead to \eref{eqn:type2compts} and
\eref{eqn:GENSOLGAUGE}.  As it turns out, the type II calculation is actually
much simpler than the the type D calculation because it uses a tetrad in which
$\tau=\tap=0$. Therefore we will work out the type D calculation in detail and
the type II result mostly follows by setting certain quantities to zero, as
indicated below.

We will need some results (and their complex conjugates) from the integration
of the type D background:
\begin{eqnarray}
\edthpt\rho &= -\deg\pi\frac{\rho}{\rhb} - \deg{\alpha}\rho -\deg\tab\rho^2,
\label{eqn:H1}\\
\rhp&=\deg{\rhp}\rhb-\half\deg \Psi\rho^2-(\edtht\deg \tab+\half \deg
\Psi)\rho\rhb-\deg \tau\deg \tab\rho^2\rhb +\deg{\tab}\deg{\pib}\rho\nonumber\\
&\pheq +\deg{\tab}\deg{\bar\alpha}\rho^2
+\half\deg{\pi}\deg{\pib}\rhb\Big(\frac{1}{\rho^2}
+\frac{1}{\rhb^2}\Big)
+\half\rhb\Big(\frac{1}{\rho}+\frac{1}{\rhb}\Big)(\edtht+\deg{\bar\alpha})
\deg{\pi},\label{eqn:rhp}\\
\tau &= -\deg{\bar\pi} - \deg{\bar\alpha}\rho + \deg\tau\rho\rhb,
\label{eqn:H2}\\
\tap &= -\deg\pi - \deg\tab\rho^2,\label{eqn:H3}\\
\Psi_2 &= \deg\Psi\rho^3.\label{eqn:H4}
\end{eqnarray}
As noted in the text, $\deg\pi\neq 0$ leads to the accelerating C-metrics,
which we include for full generality.  Henceforth the corresponding quantities
in type II spacetimes can be obtained by setting
$\deg{\tau}=\deg{\pi}=\deg{\alpha}\Rightarrow0$ and
$\deg\Psi\Rightarrow\deg{\Psi_2}$\footnote{This arises from the fact that in
type D spacetimes there is only one non-vanishing Weyl scalar, $\Psi_2$.  In
type II spacetimes, however, both $\Psi_3$ and $\Psi_4$ are in general also
nonzero. Though we do not refer to any of the other Weyl scalars in this work,
we would like maintain agreement with the standard conventions.} in the type D
result.  Thus, in type II spacetimes we have
\begin{eqnarray}
\rhp &= \deg{\rhp}\rhb -\half\deg{\Psi_2}(\rho^2+\rho\rhb),
\label{eqn:T21}\\
\tau &=0,\label{eqn:T22}\\
\tap &=0,\label{eqn:T23}\\
\Psi_2 &= \deg{\Psi_2}\rho^3,\label{eqn:T24}
\end{eqnarray}
the equation for $\edthpt\rho$ not following from the limiting process
mentioned above.  Note that the quantity $\edthpt\rho$ is never used in any of
the integrations we perform in the type II background spacetime.
We will also need the definitions of the new operators:
\begin{eqnarray}
\thpt &= \thp - \tab\edtht - \tau\edthpt +
\tau\tab(\frac{p}{\rhb}+\frac{q}{\rho}) +
\half(\frac{p\Psi_2}{\rho}+\frac{q{\bar\Psi}_2}{\rhb}),\label{eqn:O1}\\
\edtht &= \frac{\edth}{\rhb} + \frac{q\tau}{\rho},\label{eqn:O2}\\
\edthpt &= \frac{\edthp}{\rho} + \frac{p\tab}{\rhb},\label{eqn:O3}
\end{eqnarray}
where $p$ and $q$ label the GHP type of the quantity being acted on.
Additionally, in sections \ref{typeII} and \ref{typeD} we make use of the
commutator
\begin{equation}
\eqalign{
[\edtht,\edthpt] &= \frac{\rhb'-\rhp}{\rho\rhb}\th +
\Big(\frac{1}{\rhb}-\frac{1}{\rho}\Big)\thpt
+ p\Big\{\frac{\rhp}{\rhb} +
\half\Psi_2\Big(\frac{1}{\rho}+\frac{1}{\rhb}\Big) +
\edtht\Big(\frac{\tab}{\rhb}\Big)\Big\}\\
&\pheq - q\Big\{\frac{\rhb'}{\rho} +
\half{\bar\Psi}_2\Big(\frac{1}{\rho}+\frac{1}{\rhb}\Big) +
\edthpt\Big(\frac{\tau}{\rho}\Big)\Big\},
}\label{eqn:edthcomm}
\end{equation}
which is valid in type D and (with $\tau=0$) type II spacetimes.

We now begin with
\begin{equation}
\th\xi_l = 0,
\end{equation}
which integrates trivially to give
\begin{equation}
\xi_l=\deg{\xi_l}\!\label{eqn:L0}.
\end{equation}
With this information in hand, we can now integrate the equation governing
$\xi_m$:
\begin{equation}
(\th+\rhb)\xi_m + (\edth+\tab')\xi_l = 0.
\end{equation}
Rewriting the $\th$ piece and using \eref{eqn:O2} with $p=1$ leads to
\begin{equation}
\frac{1}{\rhb}\th(\rhb\xi_m) + \tab'\xi_l
+\rhb\edtht\xi_l-\frac{\rhb\tau}{\rho}\xi_l = 0,
\end{equation}
which, after substituting \eref{eqn:H2}, the complex conjugate of \eref{eqn:H3}
and \eref{eqn:L0} along with some rearranging, yields
\begin{equation}
\th(\rhb\xi_m) = -\deg{\bar\pi}\deg{\xi_l}\Big(\frac{\rhb^2}{\rho}-\rhb\Big) +
2\deg\tau\deg{\xi_l}\rhb^3 -
\rhb^2(\edtht+\deg{\bar\alpha})\deg{\xi_l}\!.
\end{equation}
Integration then gives us
\begin{equation}
\xi_m = \deg{\xi_m}\frac{1}{\rhb} -
         \deg{\bar\pi}\deg{\xi_l}\frac{1}{\rho} +
         \deg\tau\deg{\xi_l}\rhb -
         (\edtht+\deg{\bar\alpha})\deg{\xi_l}\!,\label{eqn:M0}
\end{equation}
and the solution for $\xi_{\mb}$ then follows from complex conjugation
\begin{equation}
\xi_{\mb} = \deg{\xi_{\mb}}\frac{1}{\rho} -
         \deg\pi \deg{\xi_l} \frac{1}{\rhb} + \deg\tab\deg{\xi_l}\rho -
         (\edthpt+\deg\alpha)\deg{\xi_l}\!.\label{eqn:MB0}
\end{equation}
Finally, we are in a position to deal with $\xi_n$, by writing
\begin{equation}
\thp\xi_l + \th\xi_n + (\tau+\tab')\xi_{\mb} + (\tab+\tap)\xi_m =0,
\end{equation}
in terms of Held's operators (\eref{eqn:H1}, \eref{eqn:H2} and
\eref{eqn:H3}) as
\begin{equation}
\eqalign{&\th\xi_n + \thpt\xi_l + \tab\edtht\xi_l + \tau\edthpt\xi_l -
\tau\tab\Big(\frac{1}{\rho}+\frac{1}{\rhb}\Big)\xi_l\\
&\pheq-\half\Big(\frac{\Psi_2}{\rho}+\frac{{\bar\Psi}_2}{\rhb}\Big)\xi_l
+ (\tau+\tab')\xi_{\mb} + (\tab+\tap)\xi_m = 0.}
\end{equation}
Substituting \eref{eqn:H2}, \eref{eqn:H3}, \eref{eqn:H4}, \eref{eqn:L0},
\eref{eqn:M0} and \eref{eqn:MB0}, rearranging terms and letting the dust settle
leads to
\begin{equation}
\eqalign{\th\xi_n &= -\thpt\deg{\xi_l} + \half\deg\Psi\deg{\xi_l}\rho^2+
\half\deg{\bar\Psi} \deg{\xi_l} \rhb^2 -
  \deg\pi\deg{\bar\pi}\deg{\xi_l}\Big(\frac{1}{\rho}+\frac{1}{\rhb}\Big)\\
  &\pheq + \deg\tau\deg\tab\deg{\xi_l}(\rho^2\rhb+\rho\rhb^2)
  -[\deg\tab\rho^2(\edtht+\deg{\bar\alpha})
  +\deg\tau\rhb^2(\edthpt+\deg\alpha)]\deg{\xi_l}\\
  &\pheq -  [\deg\pi(\edtht+\deg{\bar\alpha})
    +\deg{\bar\pi}(\edthpt+\deg\alpha)]\deg{\xi_l}
  +2\deg\pi\deg{\xi_m}\frac{1}{\rhb} +
  2\deg{\bar\pi}\deg{\xi_{\mb}}\frac{1}{\rho}\\
  &\pheq + \deg\tau\deg{\xi_{\mb}}\Big(\frac{\rhb^2}{\rho}-\rhb\Big) +
  \deg\tab\deg{\xi_m}\Big(\frac{\rho^2}{\rhb}-\rho\Big)
  + \deg\alpha\deg{\xi_m}
  + \deg{\bar\alpha}\deg{\xi_{\mb}}.}
\end{equation}
Integration then results in
\begin{equation}
\eqalign{\xi_n &= \deg{\xi_n} + \half\deg{\Psi} \deg{\xi_l} \rho+
\half\deg{\bar\Psi} \deg{\xi_l} \rhb +
           \deg\tau \deg\tab \deg{\xi_l} \rho\rhb +
           \half\deg\pi \deg{\bar\pi}
            \deg{\xi_l}\Big(\frac{1}{\rho^2}+\frac{1}{\rhb^2}\Big)\\
            &\pheq + \Big[\frac{\deg\pi}{\rho} (\edtht + \deg{\bar\alpha})
            +\frac{\deg{\bar\pi}}{\rhb} (\edthpt+\deg\alpha )\Big] \deg{\xi_l}
+
           \half\Big(\frac{1}{\rho}+\frac{1}{\rhb}\Big)\thpt\deg{\xi_l}\\
	    &\pheq -[\deg\tab \rho(\edtht+\deg{\bar\alpha} ) +
            \deg\tau \rhb(\edthpt+\deg\alpha )]\deg{\xi_l}
	     +\deg\tab \deg{\xi_m}\frac{\rho}{\rhb}
            +\deg\tau \deg{\xi_{\mb}} \frac{\rhb}{\rho}\\
            &\pheq-\deg\pi \deg{\xi_m} \frac{1}{\rhb^2}
            -\deg{\bar\pi} \deg{\xi_{\mb}} \frac{1}{\rho^2}
	    -\deg\alpha\deg{\xi_m}\frac{1}{\rhb}
           -\deg{\bar\alpha}\deg{\xi_{\mb}}\frac{1}{\rho},}
\end{equation}
and our task is complete.

\section{Perturbed Einstein Tensor in Schwarzschild spacetime}\label{Schwarz}

We list the components of the Einstein tensor expressed in terms of the metric
perturbations in the gauge $l^{a}h_{ab}=0$. The independent components of the
even parity metric perturbations in this gauge are, using \eref{eqn:evenIRG}:
$H(r)\equiv H_0(r)$,  $h(r)\equiv {}^eh_0(r)$, $K(r)$ and $G(r)$. The trace is
given by $T(r)=-2K(r)+\ell(\ell+1)G(r)$.  The components of the Einstein tensor
$E_1 ... E_7$ are calculated to be:

\begin{equation}
\fl \eqalign{E_{1}(r)&=
-\frac{(r-2M)}{2r}T''(r)-\frac{(3r-5M)}{2r^{2}}T'(r)
-\frac{\left[\ell(\ell+1)-2\right]}{2r^{2}}K(r)\\\bs
&\pheq
-\frac{\ell(\ell+1)}{r^{2}}h'(r)-
\frac{\ell(\ell+1)(r-3M)}{(r-2M)r^{3}}h(r)\\\bs
&\pheq
-\frac{(r-2M)}{r^{2}}H'(r)-\frac{\left[\ell(\ell+1)+2\right]}{2r^{2}}H(r),}
\end{equation}
\begin{equation}
\fl \eqalign{
E_{2}(r)&=
\frac{i\omega}{2}T'(r)+\frac{i\omega(r-3M)}{2r(r-2M)}T(r)
+\frac{\left[\ell(\ell+1)+2ir\omega\right]}{2r^{2}}H(r)\\\bs
&\pheq +\frac{\ell(\ell+1)}{2r^{2}}h'(r)+\frac{\ell(\ell+1)(i\omega
r^{2}-2M)}{2(r-2M)r^{3}}h(r),}
\end{equation}
\begin{equation}
\fl \eqalign{
E_{3}(r)&=
\frac{(r-M)}{2r^{2}}T'(r)+\frac{r\omega^2}{2(r-2M)}T(r)+\frac{\ell(\ell+1)
(r-M-i\omega r^2)}{r^{3}(r-2M)}h(r)\\\bs
&\pheq
+\frac{(r-2M)}{r^{2}}H'(r)-\frac{2ir\omega}{r^{2}}H(r)-
\frac{\left[\ell(\ell+1)-2\right]}{2r^{2}}\left[H(r)-K(r)\right],}
\end{equation}
\begin{equation}
\fl \eqalign{
E_{4}(r)&=
\frac{(r-2M)}{2r}h''(r)-\frac{i\omega}{2}h'(r)+\frac{\left[2M(r-2M)-i\omega
r^{2}(r-3M)\right]}{r^{3}(r-2M)}h(r)\\\bs
&\pheq +\frac{(r-2M)}{2r}H'(r)+\frac{(2M-i\omega
r^{2})}{2r^{2}}H(r)-\frac{i\omega}{2}\left[K(r)-G(r)\right],}
\end{equation}
\begin{equation}
\fl \eqalign{
E_{5}(r)&=
-\frac{i\omega r}{2(r-2M)}h'(r)-\frac{\left[2(r-2M)(1-i\omega
r)+r^{3}\omega^2\right]}{2r(r-2M)^{2}}h(r)\\\bs
&\pheq -\frac{1}{2}H'(r)-\frac{(2M-i\omega
r^{2})}{2r(r-2M)}H(r)+\frac{1}{2}\left[K'(r)-G'(r)\right],}
\end{equation}
\begin{equation}
\fl \eqalign{
E_{6}(r)&=
\frac{(r-2M)}{2r}H''(r)+\frac{(1-i\omega
r)}{r}H'(r)-\frac{i\omega(2r-2M-i\omega r^{2})}{2r(r-2M)}H(r)\\\bs
&\pheq
+\ell(\ell+1)\Bigl[\frac{(r-2M)}{2r}G''(r)+\frac{(r-M)}{r^{2}}G'(r)+
\frac{r\omega^2}{2(r-2M)}G(r)\Bigr] \\\bs
&\pheq
-\Bigl[\frac{(r-2M)}{2r}K''(r)+\frac{(r-M)}{r^{2}}K'(r)+
\frac{r\omega^2}{2(r-2M)}K(r)\Bigr] \\\bs
&\pheq +\frac{\ell(\ell+1)}{r^{2}}h'(r)
-\frac{i\omega\ell(\ell+1)}{r(r-2M)}h(r),}
\end{equation}
\begin{equation}
\fl \eqalign{
E_{7}&=
\frac{(r-2M)}{2r}G''(r)+\frac{(r-M)}{r^{2}}G'(r)+
\frac{r\omega^2}{2(r-2M)}G(r)\\\bs
&\pheq +\frac{1}{r^{2}}h'(r)-\frac{i\omega}{r(r-2M)}h(r).}
\end{equation}

The independent components of the odd parity perturbations in this gauge are,
using \eref{eqn:oddIRG}: $h(r)\equiv {}^oh_0(r)$ and $H(r)\equiv h_2(r)$. The
components of Einstein tensor $F_1 ... F_3$ are calculated to be

\begin{equation}
\fl \eqalign{
F_{1}(r)&=
\frac{r-2M}{2r}h''(r)-\frac{i\omega}{2}h'(r)+\frac{i\omega\left[\ell(\ell+1)-
2\right]}{4r^{2}}H(r)\\\bs
&\pheq -\frac{i\omega r^{3}+\left[\ell(\ell+1)/2-i\omega
3M\right]r^{2}-\left[\ell(\ell+1)+2\right]rM+4M^{2}}{r^{3}(r-2M)}h(r),}
\end{equation}
\begin{equation}
\fl \eqalign{
F_{2}(r)&=
-\frac{\left[\ell(\ell+1)-2\right]}{4r^{2}}H'(r)-\frac{i\omega
r}{2(r-2M)}h'(r)+\frac{\left[\ell(\ell+1)-2\right]}{2r^{3}}H(r)\\\bs
&\pheq -\frac{r^{3}\omega^2/2-(r-2M)\left[i\omega
r-1+\ell(\ell+1)/2\right]}{r(r-2M)^{2}}h(r),}
\end{equation}
\begin{equation}
\fl \eqalign{
F_{3}(r)&=
\frac{(r-2M)}{2r}H''(r)-\frac{(r-3M)}{r^{2}}H'(r)-h'(r)\\\bs
&\pheq +\frac{i\omega
r}{(r-2M)}h(r)+\frac{(r^{4}\omega^{2}/2+r^{2}-6Mr+8M^{2})}{r^{3}(r-2M)}H(r).}
\end{equation}

\section*{References}
\bibliography{rad_rough}
\bibliographystyle{amsplain}
\end{document}